\documentclass[letterpaper,final]{iopart}
\usepackage{epsfig}
\begin{document}

\title[Impurity effects on the band structure of one-dimensional photonic crystals]
{Impurity effects on the band structure of one-dimensional photonic crystals: Experiment and theory}

\author{G.A. Luna-Acosta$^{1\dagger}$ and H. Schanze$^2$, U. Kuhl$^2$, and H.-J. St\"{o}ckmann$^2$}
\address{$^1$ Instituto de Fisica, BUAP Apartado Postal J-48, 72570 Puebla, M\'exico.}
\address{$^2$ Fachbereich Physik der Philipps-Universit\"{a}t Marburg, Renthof 5, D-35032, Germany.}
\ead{$^\dagger$ gluna@sirio.ifuap.buap.mx}
\date{\today}

\begin{abstract}
We study the effects of single impurities on the transmission in
microwave realizations of the photonic Kronig-Penney model,
consisting of arrays of Teflon pieces alternating with air spacings
in a microwave guide. As only the first propagating mode is
considered, the system is essentially one dimensional obeying the
Helmholtz equation. We derive analytical closed form expressions
from which the band structure, frequency of defect modes, and band
profiles can be determined. These agree very well with experimental
data for all types of single defects considered (e.\,g.\
interstitial, substitutional) and shows that our experimental set-up
serves to explore some of the phenomena occurring in more
sophisticated experiments. Conversely, based on the understanding
provided by our formulas, information about the unknown impurity can
be determined by simply observing certain features in the
experimental data for the transmission. Further, our results are
directly applicable to the closely related quantum 1D Kronig-Penney
model.
\end{abstract}
\maketitle

\tableofcontents

\section{Introduction}

Research on photonic crystals, theoretical and experimental, has
been sustained at a high intensity for several years, ever since the
publications of Yablonovitch \cite{Yab87} and John \cite{Joh87} in
1987. One interest for this research is the potentially very high
number of applications in optoelectronics (see, e.\,g., Chapter~VII
of~\cite{Joa95}). A substantial part of these studies
concerns the study of impurities or defects in one, two, or three
dimensional (1D, 2D or 3D) photonic crystals.

Impurities or defects in photonic crystals may sometimes be unwanted
but may also be extremely useful. For example, impurity states lying
in a complete photonic band gap can be used for a waveguide and thus
be an essential part of optical devices \cite{Joh02}. By introducing
defects periodically in a perfect photonic array, coupled cavity
waveguides are formed. The coupling of the cavity modes create
impurity bands which have potential applications for the design of
high-efficiency waveguides and waveguide bends \cite{Bay00,XuS04}.
It is important to mention that defects can be introduced in a
controlled manner in photonic array experiments, see, e.\,g.~\cite{Qi04}.
There are promising theoretical results
\cite{Cha05} that point defects, in particular a substitutional
defect in 3D crystals formed by a lattice of air spheres on a
silicon background, may be used as micro-cavity for localizing light
at a given point. Whence we appreciate the technological, as well as
academic, importance of understanding the effects of various types
of defects or impurities.

There are several calculational methods for the treatment of impurities in
photonic crystals. These are based on plane wave expansion of the
fields \cite{Joh01}; finite-difference time-domain algorithms
\cite{Kun93,Qi04}; variational methods \cite{Cha05}; R-matrix
methods \cite{Els96}; transfer-matrix methods \cite{Pen92},
combined, if necessary, with finite element methods; super-cell
\cite{Kha98}, Green-function, and tight-binding methods (see,
e.\,g.\ Ref.~\cite{Bay00}). Eigenfrequencies and
eigenfunctions of defects can also be calculated via the method of
Sakoda et~al.~\cite{Sak01} based on the numerical simulation of the
excitation process of the defect mode by a virtual oscillation
dipole moment, in conjunction with the finite-time domain algorithm.

On the other hand, for {\it electronic} periodic systems the
calculation of impurity states dates back to the mid 50's with
the introduction of effective mass theory \cite{Kit54,Lut55}.
Useful modern methods for the calculation of impurity levels are the
super-cell tight-binding methods \cite{Noz00}, which are applicable
to shallow, deep and intermediate impurity levels \cite{Men99}.

In this work we investigate experimentally and theoretically the
effect of single defects or impurities in the transmission of the
electromagnetic field through arrays formed by Teflon pieces
alternating with air spacings. A closely related system to ours has
been studied by Pradhan et~al.~\cite{Pra99} who looked for effects
of isolated impurities in a system formed by an array of coaxial
connectors.

It is well known that point defects can produce localized states in
the gaps \cite{Joa95,Sak01}. The first experimental observation in
photonics was made by McCall et.\,al.\ \cite{McC91}. Can we observe
these in our experiment? How else do impurities manifest themselves
in the transmission curves? In this paper we investigate these
questions for various types of single defects. We compare
experimental results with those obtained by the transfer matrix
calculations, equation~(\ref{eq:eq13}), and point out the most prominent
and typical transport features.

One purpose of this paper is to show that our experimental set up
can be used as a test-bed to study some of the phenomena occurring
in more sophisticated (and expensive) photonic or electronic
experimental arrangements. The second purpose is to show that our
analytical expressions, derived in Section IV for the photonic
Kronig-Penney model with single defects, are very helpful in
elucidating the various features in the band structure; in
particular, the frequencies of the defect modes and band profiles
observed in the experiments.

We remark that since our system is described by the one dimensional
Helmholtz equation our results are directly applicable also to
one-dimensional semiconductor (electronic) crystals formed, e.\,g., by
sequences of quantum dots. Effects of irregularities such as an
additional scatterer or a displaced quantum dot from its regular
position have been the focus of many investigations, especially
since these were observed experimentally by Kouwenhoven
{\it{et~al.}}~\cite{Kou90} in the case of electronic transport in
heterostructures and by McCall~\cite{McC91} in the case of photonic
crystals.

\section{Experimental set-up and the model}

\begin{figure}
\hspace*{2.5cm}\includegraphics[height=4cm]{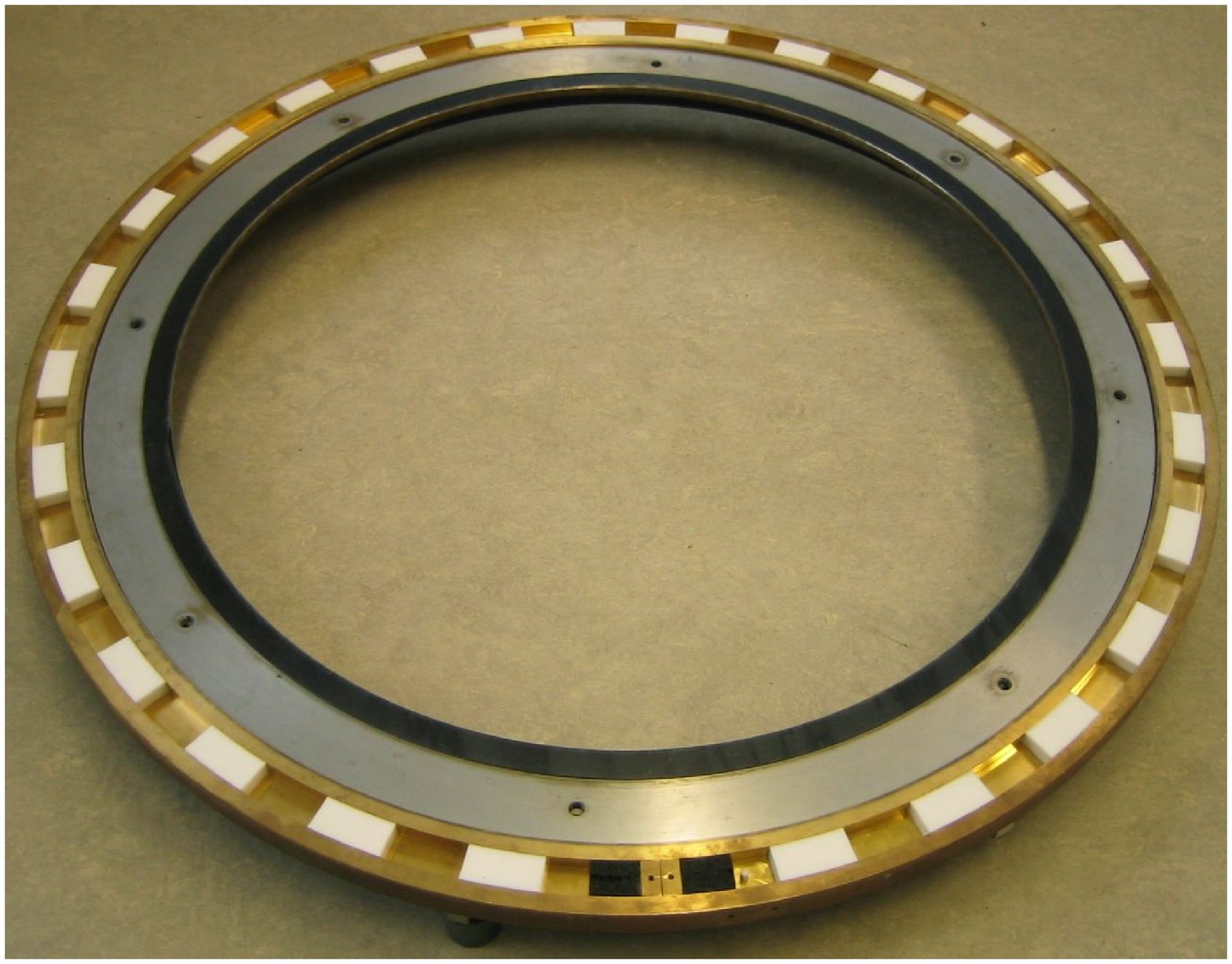}
\includegraphics[height=4cm]{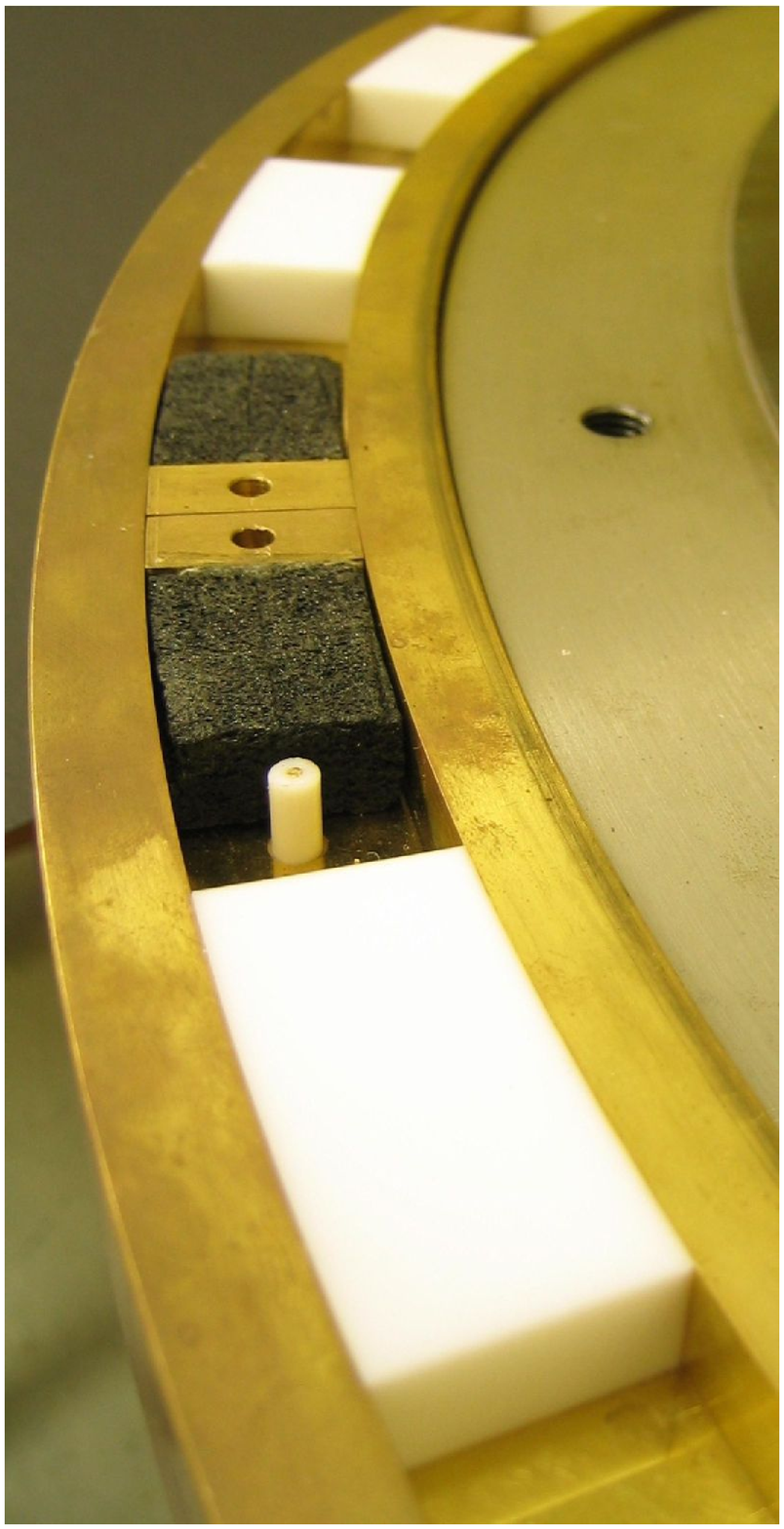}
\caption{\label{fig:fig1} Experimental set-up. Left:
Overview of the wave guide. Right: Enlargement of the part close to
the electric dipole antenna and the two carbon absorbers. The top
brass plate covering the wave guide is not shown. The wave guide has
a total diameter of 78\,cm, a depth $A$ = 1\,cm and a width $B$ =
2\,cm. The frequency range of the first propagating mode is from
$7.5$ to $15$ GHz. The length of the Teflon pieces shown is $d'$ =
4\,cm and their index of refraction is $n=\sqrt{2.08}$. In this
picture, the spacing between all Teflon pieces is $d$ = 4\,cm.}
\end{figure}

The experimental set-up is shown in figure~\ref{fig:fig1}. It
consists of a brass ring where a microwave guide is cut out, in
which several Teflon pieces, two carbon absorbers, and an antenna
are inserted. Another antenna is fixed on the top plate (not shown
in the figure) covering the waveguide. The antennas are connected to
a network analyzer that allows the measurement of all matrix
elements of the scattering matrix. The absorbers are used to
eliminate reflection at the ends of the array and hence mimic a 1D
scattering system with open ends. This circular wave guide has been
used, with metallic screws instead of Teflon pieces, to study the
microwave realization of the Hofstadter butterfly \cite{Kuh98b} and
transport properties of 1D on-site correlated disorder potentials \cite{Kuh00a}.

\begin{figure}
\hspace*{2.5cm}\includegraphics[width=7cm, height=4cm]{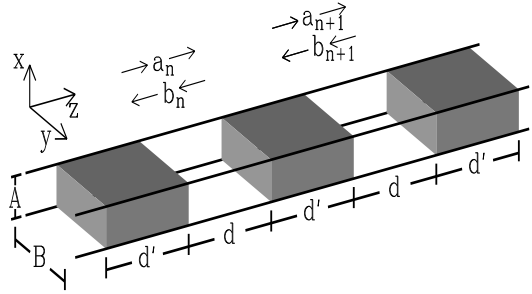}
\caption{\label{fig:fig2} The photonic Kronig-Penney model. The gray
blocks are the idealized Teflon pieces of length $d'$, the air
spacing is $d$. $a_n$ and $b_n$ are the amplitudes of the wave
function in the air spacing to the right of the $n$th Teflon piece.}
\end{figure}

In our experiment, the cut-off frequency for the lowest mode is
$f_{\rm min}= c/2B \approx 7.5$\,GHz and the second mode opens at $15$
GHz. All results presented in this paper are restricted to the
regime of the first propagating mode. Thus the system is effectively
one-dimensional. Since the perimeter of the ring ($\approx 234$\,cm)
is much larger than the maximum wavelength used in our
experiments ($2<\lambda<4$\,cm), our theoretical model assumes, as
an approximation, a linear set-up (see figure~\ref{fig:fig2}).

For the lowest TE mode ($E_z=0$) at $f<15$\,GHz the $y$ component
of the electric field $E_y$ is also zero and the $x$ component is
\begin{displaymath}\label{eq:eq0}
E_x(y,z,t)=E_o \sin\left(\frac{\pi y}{B}\right)\Phi(z)\exp(i2\pi
ft), \nonumber
\end{displaymath}
The wave function $\Phi(z)$ obeys the Helmholtz equation
\begin{equation}\label{eq:eq1}
\left(\frac{d^2}{d z^2}+k_r^2\right) \Phi(z)=0,
\end{equation}
where
\begin{equation}\label{eq:eq2}
k_r=\sqrt{\frac{(2\pi f)^2}{c^2}n_r^2-\frac{\pi^2}{B^2}}
\end{equation}
and $n_r=\sqrt{\mu\epsilon}$ is the position-dependent index of
refraction. In the case the Teflon pieces are periodically spaced
our system is the electromagnetic counterpart of the quantum 1D
Kronig-Penney model.

We use the transfer matrix approach to calculate the transmission.
Since both antennas are placed in the air, our transfer matrix
should connect the wave functions amplitudes from air to air.
Referring to figure~\ref{fig:fig2}, the transfer matrix $Q$ for a single cell
connects the amplitudes $(a_n, b_n)$ and $(a_{n+1}, b_{n+1})$:
\begin{equation}\label{eq:eq3}
\left(\begin{array}{c} a_{n+1}\\ b_{n+1}
\end{array}\right)
=Q\left(\begin{array}{c} a_n\\ b_n
\end{array}\right), \,\,\, \rm{with} \, Q =\left(\begin{array}{cc} Q_{11} & Q_{12}
\\ Q_{21} & Q_{22}\end{array}\right),
\end{equation}
and $k$ ($k'$) denotes the wave vector in air (Teflon pieces)
(c.\,f.~equation~(\ref{eq:eq2})). The transfer matrix for a single cell are
\begin{eqnarray}\label{eq:eq4}
 Q_{11}&=& \left[\cos(k'd')+i\alpha_+\sin(k'd')\right]e^{ikd}\\
 \label{eq:eq5}
 Q_{12}&=&-i \alpha_-\sin(k'd'),
\end{eqnarray}
where
\begin{equation}\label{eq:eq6}
\alpha_{\pm}=\frac{k^2 \pm k'^2}{2kk'}.
\end{equation}

Due to preservation of flow and time reversibility it yields
$Q_{11}= Q_{22}^*$, $Q_{12}=Q_{21}^*$ and $\det(Q)=1$.
The elements $Q_{ij}$ are the same as in the quantum 1D model of a
square potential well, but with $k=\sqrt{2\mu E}$, $k'= \sqrt{2
\mu(E -V_0)}$ and $V_0>0$. From (2) $k'>k$; thus, the Teflon piece
acts as a well ($V_0<0$) in quantum mechanics, except that in the
photonic array the ``depth'' increases with frequency.

The microwave vector network analyzers measure the full scattering
matrix $S$ defined by
\begin{equation}\label{eq:eq7}
\left(\begin{array}{c} a_{n+1}\\ b_{n}
\end{array}\right)
=S\left(\begin{array}{c} a_n\\ b_{n+1}
\end{array}\right).
\end{equation}
In terms of the $Q$ matrix elements, the $S$ matrix reads
\begin{equation}\label{eq:eq8}
S= \left(\begin{array}{cc} -\frac{\displaystyle
Q_{21}}{\displaystyle Q_{22}} & \frac{\displaystyle 1}{\displaystyle
Q_{22}}\\[1.5ex]
 \frac{\displaystyle 1}{\displaystyle Q_{22}}& \frac{\displaystyle Q_{12}}{\displaystyle Q_{22}}
\end{array}\right).
\end{equation}
The transmission $T$ through a single cell is given by
\begin{equation}\label{eq:eq9}
T=|S_{12}|^2=\frac{1}{|Q_{22}|^2}=\frac{1}{1+|Q_{12}|^2},
\end{equation}
with $Q_{12}$ and $Q_{22}$ given by Eqs.~(4) and (5) for a single
cell.

Expression (9) together with equation~(\ref{eq:eq5}) shows immediately that a single
Teflon piece becomes completely transparent at frequencies obeying
the relation
\begin{equation}\label{eq:eq10}
k'd'=m\pi, \quad \quad m =1,2,3,\dots
\end{equation}
We shall refer to these frequencies as {\it 1-Teflon resonances}.

For an array of N equally spaced cells without any impurities or
defects, we need to evaluate $Q$ to the $Nth$ power; a numerical
process that can be easily performed. However, it is more
illustrative to use the Cayley-Hamilton theorem of linear algebra to
get \cite{Gri01,TBP,Spr93}
\begin{equation}\label{eq:eq11}
T=\frac{\displaystyle 1}{\displaystyle 1+|Q_{12}^N|^2}=\frac{\displaystyle 1}{1+\left|Q_{12}\frac{\displaystyle \sin(N\theta)}{\displaystyle \sin(\theta)}\right|^2
},
\end{equation}
where $\theta$ is the Bloch phase corresponding to the infinitely
periodic array:
\begin{equation}\label{eq:eq12}
\cos \theta=\Re Q_{11} =\cos(k'd')\cos(kd)-\alpha_+\sin(k'd')\sin(kd).
\end{equation}

Note that $T=1$ not only when $Q_{12} = 0$ (i.\,e., at the 1-Teflon
resonances) but also whenever $N\theta= \pm n\pi, n=1,2,\dots,N-1$. The
latter one gives rise to $N-1$ peaks with $T=1$ in each transmission
band since $\theta$ shifts through $\pi$
\cite{Gri01,TBP,Spr93,Bar99f}. According to the above condition, the
system of $N=2$ Teflon pieces separated by air becomes transparent
when $\theta=\pi/2$, which in turn implies $\rm{Tr} Q=0$. We call
these the {\it 2-cells resonances} emphasizing that it is not just
two Teflon pieces next to each other but separated by an air
spacing. A little thought reveals that the $N-1$ oscillations are
centered around the 2-cells resonance frequency (see also
\cite{Spr93}).

\begin{figure}
\hspace*{2.5cm}\includegraphics[width=0.7\columnwidth]{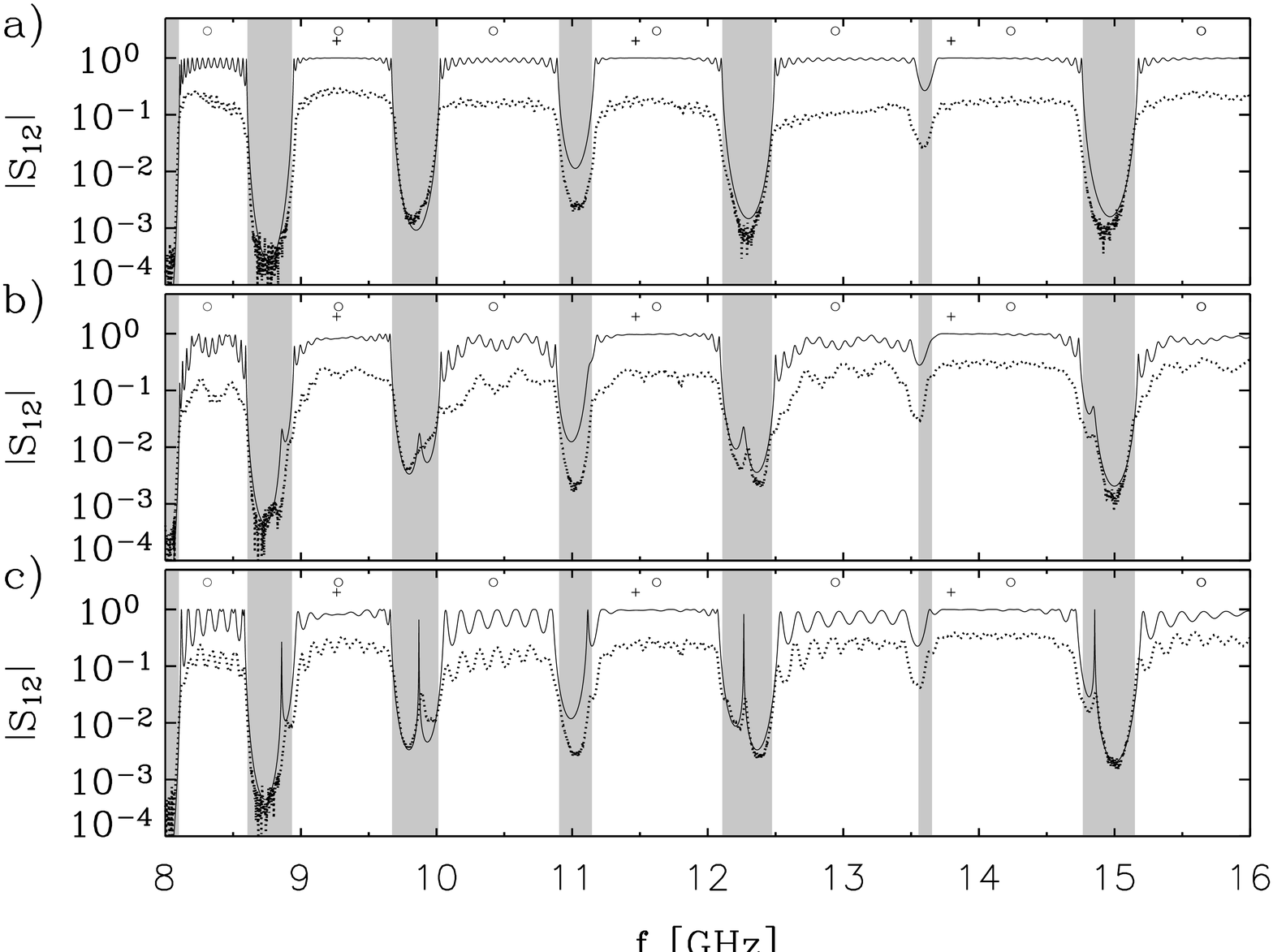}
\caption{\label{fig:fig3} Interstitial impurities for a 16 cells
array with $d=d'=4$\,cm. Dotted (solid) lines are the experimental
(theoretical) curves. The shaded regions mark the forbidden gaps,
given by the condition $\xi^2<1$ (a) pure (impurity-free) array; (b)
small Teflon piece (3.16\,cm) inserted in the center between the 3rd
and 4th regular Teflon pieces, (c) small Teflon piece (3.16\,cm) inserted
in the center between the 8th and 9th regular Teflon pieces. The
crosses (circles) mark the frequency values for the 1-Teflon-resonances
(2-cells resonance), see text.}
\end{figure}

Luna-Acosta \etal \cite{TBP} treat the case of the regular Kronig-Penney model
(i.\,e., periodic and no defects) where it is shown that
equation~(\ref{eq:eq11}) reproduces most details of the experimental
transmission data as a function of frequency for all kinds of
periodic arrays considered. Since we are concerned here with the
effects of the impurities, we show for comparison the theoretical
and experimental results for the transmission amplitude $|S_{12}|$
for an array 16 equally spaced cells, with Teflon pieces of width
$d'$=4.0\,cm and air spacing  of width $d$= 4.0\,cm, see
figure~\ref{fig:fig3}(a).

For the experimental and theoretical curves to agree as well, it was
necessary to define some effective length for the length of the
Teflon pieces and the air spacings. That is, since the actual
waveguide is circular, the Teflon pieces and air spacings are not
rectangular pieces but slightly curved with the larger side being
exactly 5 percent larger than the shorter side. (The 4\,cm Teflon
pieces and the air spacings referred to above actually mean that the
shorter side is 4\,cm whereas the larger one is 4.2\,cm.) It turned
out that the best fit could be obtained with an effective length of
4.08\,cm, i.e, a 2 percent larger than its shorter side. We
emphasize  that this is the only fitting parameter in our
calculations and moreover it is the same for all calculations
presented here. Throughout the paper, whenever we quote the width of
a Teflon piece or an air spacing we mean the shorter side of it and
in the numerical calculations we use their corresponding effective
length.

We remark that our model does not consider the absorption of the signal.
However, comparison of the experimental and theoretical data shows
that the band structure (gaps and band profiles) is not affected by
the absorption except for the attenuation of the transmission, which
is about constant throughout the frequency range (the experimental
transmission is about five times weaker). Thus, absorption in our
experiment, does not destroy coherent phenomena like
the band structure (see also \cite{Kuh00a}).

In figure~\ref{fig:fig3} and all subsequent transmission plots the frequency values
of the 1-Teflon and 2-cells resonances are marked by crosses and
circles, respectively. Different types of bands are formed depending
on the position of the Teflon resonances relative to the 2-cells
resonances \cite{TBP}. In these and all forthcoming transmission
plots we mark with shaded strips the gap regions, defined by the
condition that the eigenvalues $\lambda_{\pm}$ of the transfer
matrix $Q$ are real. Since $\lambda_{\pm}=\xi \pm \sqrt{\xi^2-1}$,
where $\xi=\Re Q_{11}$, forbidden gaps occur whenever $\xi^2>1$.

Given the good agreement between experimental transmission data and
the photonic Kronig-Penney model, implied by equation~(\ref{eq:eq11}), we proceed
to discuss the effects of impurities in the photonic Kronig-Penney
model.

\section{Impurity effects in the transmission}

\begin{figure}
\hspace*{2.5cm}\includegraphics[width=.7\columnwidth]{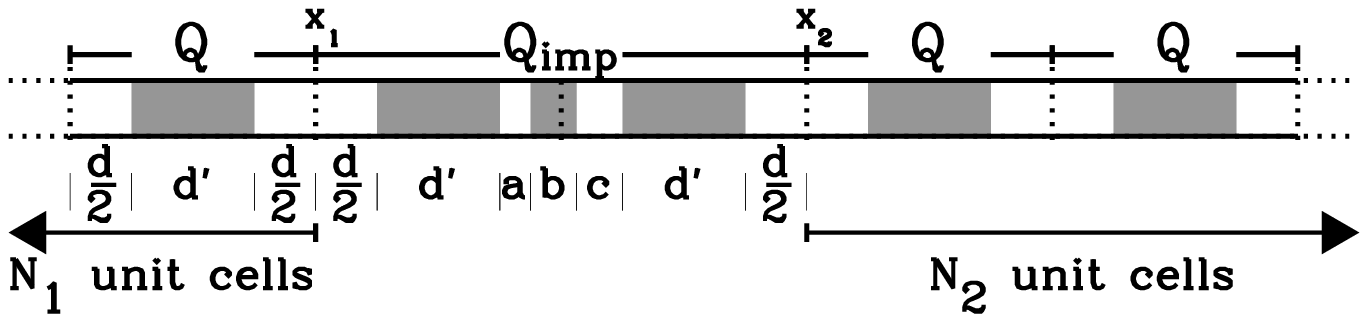}\\
\hspace*{2.5cm}\includegraphics[width=.7\columnwidth]{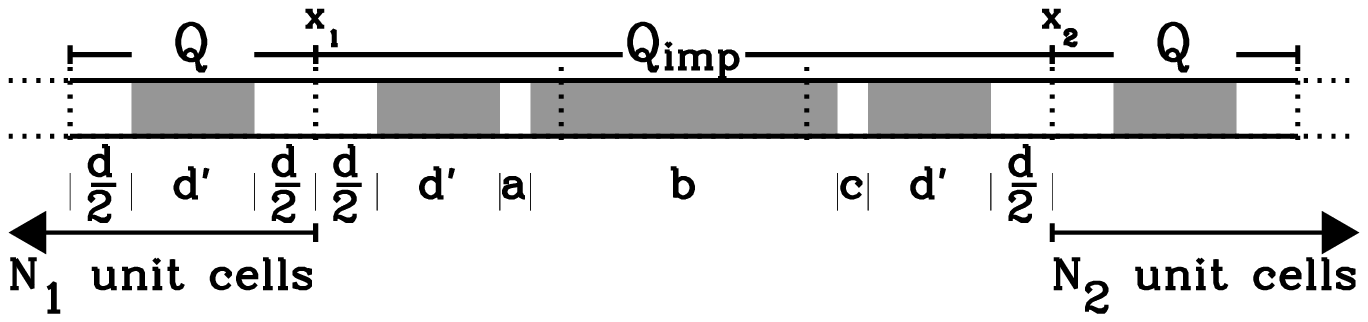}
\caption{\label{fig:fig4} Sketch of impurity arrangements. In the
upper part the interstitial, in the lower part a substitutional
impurity is shown. In the case of the substitutional $N_1+N_2= N-3$,
whereas in the case of the interstitial $N_1+N_2= N-2$, where $N$ is
the total number of unit cells of the perfect crystal.}
\end{figure}

Different types of defects or impurities can be realized in our
experimental set-up with just air and Teflon segments and can be
represented by one of the two general sketches shown in figure~\ref{fig:fig4}. The
upper sketch illustrates a general interstitial impurity: an extra
piece of Teflon of width $b$ placed somewhere in the air spacing
between two pieces of Teflon. The rest of the arrangement remains
unaltered; the perpendicular dotted lines mark the boundaries of the
unit cell along the {\it pure} crystal. By definition of point
defect, the interstitial impurity breaks the periodicity only
locally. Hence, the length of the extra piece of Teflon should be
less than or equal to the air spacing. If the width $b$ is larger
than this, then the boundaries of the regular cells would be
displaced from their original position and consequently periodicity
would be globally broken (it becomes an extended defect, e.\,g., a
topological defect \cite{Gum03,Mor03}). In this paper we do not
consider topological defects.

As shown in the upper sketch there are $N_1$ unit cells periodically
arranged to the left of point $x_1$, each described by the matrix
$Q$, then periodicity is interrupted by the impurity at point $x_1$
and again recovered at point $x_2$. To the right of point $x_2$ there
are $N_2$ cells periodically arranged. The total transfer matrix
going from left end of the array to right end of the arrays can be
written as
\begin{equation}\label{eq:eq13}
 Q_{\rm tot}=Q^{N_2}Q_{\rm imp}Q^{N_1},
\end{equation}
where $Q_{\rm imp}$ is the transfer matrix connecting the amplitudes of
the wave function at $x_1$ with those at point $x_2$. Specifically,
\begin{equation}\label{eq:eq14}
Q_{\rm imp}= D_{d/2}MD_cM_bD_aMD_{d/2},
\end{equation}
where $D_i$ ($i =d/2, a$, and $c$), is the transfer matrix
corresponding to an air spacing, of length $i$. $M$ is the transfer
matrix for the regular size Teflon piece and $M_b$ that of the
Teflon piece of length $b$.Note that $N_1+N_2= N-2$, where $N$ is
the total number of unit cells (and Teflon pieces) of the perfect
crystal.

The general case of a single substitutional impurity is shown in the
lower sketch of figure~\ref{fig:fig4}. The substitutional impurity has a length
$b$, shown here longer than the regular size Teflon piece, but it
could be of any length and/or of a different dielectric material.
Quick inspection shows that the form of the impurity matrix is the
same, but the values of the lengths $a$, $b$, and $c$ are different
and now $N_1+N_2= N-3$.

Thus, we have a straight-forward {\it numerical} scheme to calculate
the transmission amplitude in the presence of an impurity: for a
given impurity arrangement we compute equation~(\ref{eq:eq13}).

In figure~\ref{fig:fig3}(b) we show the case when an extra Teflon piece of smaller
length (than the length of the default Teflon piece) is inserted in
the center of the air spacing between the 3rd and 4th regular Teflon
pieces, thus $N_1=2, N_2=12)$. Figure~\ref{fig:fig3}(c) shows the case of the same
type of impurity but located in the air spacing between the 8th and
9th regular Teflon pieces, thus $N_1=N_2=7$. Notice that the good
agreement between the transfer matrix calculations and main
features of the experimental data.

Inspection of the plots of figure~\ref{fig:fig3} reveals that the impurity
affects the transmission in two ways: i) The bands develop different
profiles (compared to the pure array, figure~\ref{fig:fig3}(a)). ii) Peaks appear
in most of the gaps in figure 3(b) coinciding in frequency with those
of figure~\ref{fig:fig3}(c). Such a coincidence is expected since the impurity in
both arrays occupies the same relative position {\it within} the
cell (in the center of the air spacing) but in a different cell.

In figure~\ref{fig:fig3}(b) the band profiles look like a superposition of slow
and fast oscillations. In figure~\ref{fig:fig3}(c) the oscillations are larger and
only of one type. In both, figure~\ref{fig:fig3}(b) and (c), bands containing
1-Teflon resonances (see the 2nd, 4th, and 6th bands) are less
affected by the impurity. Note also that the intensity of the peaks
is greater when the impurity is placed near the center of the array,
figure~\ref{fig:fig3}(c), than when it is placed near the end of the array, figure~\ref{fig:fig3}(b).
This feature was invariably found in all our experiments.

\begin{figure}
\hspace*{2.5cm}\includegraphics[width=0.7\columnwidth]{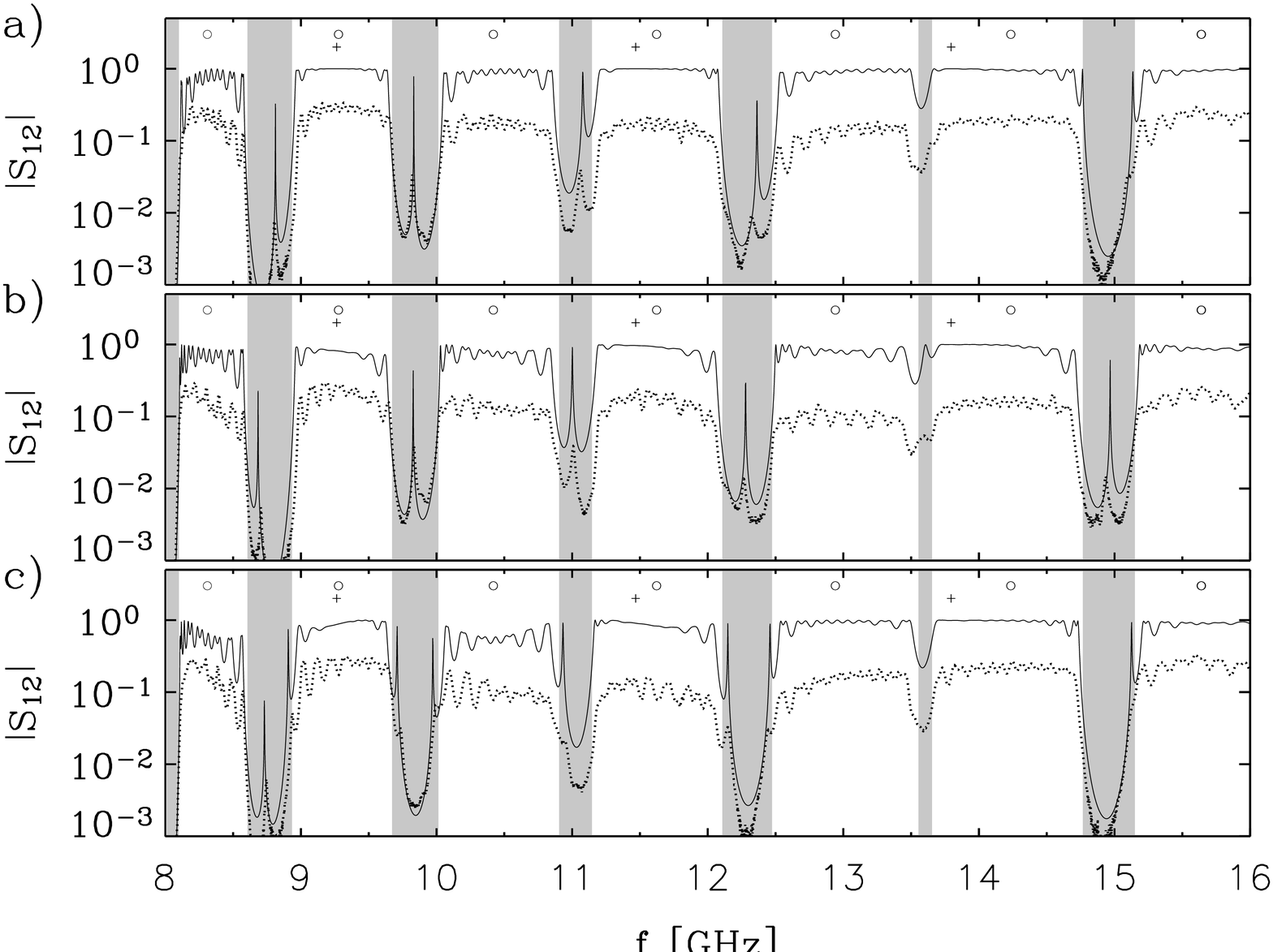}
\caption{\label{fig:fig5} Three types of substitutional impurities
in an 16 cell array ($d=d'=4$\,cm). Dotted (solid) lines are the
experimental (theoretical) curves. In (a) the 8th Teflon piece was
removed corresponding to a vacancy. In (b), (c) it was replaced by a
smaller (larger) size Teflon of length $d'=3.16$\,cm
($d'=6.32$\,cm), respectively.}
\end{figure}

Figure~\ref{fig:fig5} shows the experimental and theoretical curves for three types
of substitutional impurities. Figure~\ref{fig:fig5}(a) pertains to the case when a
Teflon is removed from the array, {\it i.\,e.,} a vacancy.
Specifically, the 8th Teflon is removed. In figure~\ref{fig:fig5}(b) the 8th
regular size Teflon piece is substituted by a smaller, whereas in (c)
it is substituted by a larger one. Note that while the defect is in
the same number of cell within the array, the band profiles, the
frequencies and intensities of the peaks are different for each case
since the defects are different.

A single point defect can produce multiple impurity levels within a
band, as has been observed, e.\,g.\ by Yablonovitch
et~al.~\cite{Yab91b} in an three dimensional photonic array with a
donor-like impurity. Note that the figure~\ref{fig:fig5}(c) shows also two spikes
in most of the gaps.

Shallow impurities in semiconductors, associated with long-range
binding, typically show up as spikes near the band edges, whereas
deep impurities typically lie near the center of the gap
\cite{Men99}. However, figure~\ref{fig:fig5} shows that a single impurity may
produce spikes near the center of the gap or near the band edges.
Thus we see that the above characterization of the impurity as
either deep or shallow is not always applicable.

Although our numerical procedure is seen to reproduce quite well the
main features of the experimental data, it does not serve to explain
their origin. For this purpose we develop a scheme in the next
section which provides us with a general understanding of the direct
relation between the number of slow and large or fast and small
oscillations and the position of the impurities in the array. We
will also show how to determine the frequency position of the
impurities.

\section{Analysis of the effects of the impurities.}

We express the transfer matrix $Q$ in terms of its diagonal
representation $\Lambda$. Inserting $Q=P\Lambda P^{-1}$ into
equation~(\ref{eq:eq14}) to get

\begin{equation}\label{eq:eq15}
Q_{\rm tot}=P\Lambda^{N_2}\tilde Q\Lambda^{N_1}P^{-1},
\end{equation}
where
\begin{equation}\label{eq:eq16}
\tilde Q=P^{-1}Q_{\rm imp}P,
\end{equation}
with
$\displaystyle P = \left(\begin{array}{cc}
\frac{Q_{12}}{\lambda_+-Q_{11}} & \frac{Q_{12}}{\lambda_--Q_{11}} \\
1 & 1 \end{array} \right),
\Lambda = \left(\begin{array}{cc}
\lambda_+ & 0 \\ 0 & \lambda_-\end{array} \right )$,
$\lambda_{\pm}= \xi \pm \sqrt{\xi^2-1}$, and $\xi=\mbox{Tr} Q/2.$

Note that $\tilde Q$ is the impurity matrix in the representation of
the diagonal basis of $Q$. This ``rotated impurity matrix'' contains
information about the coupling of the impurity to the host material
through the elements of the transformation matrix $P$.

From the definition of $Q_{\rm tot}$ and using the representation of
$P$ given above we get after some algebra useful expressions for the
matrix elements of $Q_{\rm tot}$:

\begin{equation}\label{eq:eq17}\fl
Q_{\rm tot,22} =\lambda_+^{N_1+N_2} \left(f_+{\tilde Q_{11}}\;+ f_+\tilde
Q_{21}\lambda_+^{{-2N_2}}-f_-\tilde
Q_{12}\lambda_+^{{-2N_1}}-f_-\tilde Q_{22}\lambda_+^{-2(N_1+N_2)}\right),
\end{equation}
where $f_\pm=(\lambda_\pm-Q_{11})/(\lambda_+-\lambda_-)$,
\begin{equation}\label{eq:eq18}\fl
\tilde Q_{21}= Q_{\rm imp,11}f_- + \frac{Q_{\rm imp,12}}{Q_{12}}(\lambda_-
-Q_{11})f_+ - \frac{Q_{\rm imp,21} Q_{12}}{\lambda_+ -Q_{11}}f_-
-Q_{\rm imp,22}f_- ,
\end{equation}
and
\begin{equation}\label{eq:eq19}\fl
\tilde Q_{22}=Q_{\rm imp,11}f_+ + \frac{Q_{\rm imp,12}}{Q_{12}}(\lambda_-
-Q_{11})f_+ - \frac{Q_{\rm imp,21} Q_{12}}{\lambda_+
-Q_{11}}f_+-Q_{\rm imp,22}f_-.
\end{equation}
The terms $\tilde Q_{11}$ and $\tilde Q_{12}$ are obtained from
$\tilde Q_{22}$ and $\tilde Q_{21}$, respectively, by exchanging
$\lambda_+$ with $\lambda_-$ everywhere. As in the case of a single
scatterer, it is also true that the complex conjugate of
$Q_{\rm tot,11}$ ($Q_{\rm tot,12}$) is $Q_{\rm tot,22}$ ($Q_{\rm tot,21}$). This property
is not obeyed by the elements of the rotated impurity matrix. On the
other hand, because equation~(\ref{eq:eq16}) is a unitary transformation,
$\mbox{Tr}(\tilde Q)=\mbox{Tr}(Q_{\rm imp})$ and $\det(\tilde
Q)=\det(Q_{\rm imp})$.

Setting $N_1+N_2=N, \tilde Q_{11}=\tilde Q_{22}$, and $\tilde
Q_{12}=\tilde Q_{21}=0$ in equation~(\ref{eq:eq16}) and using $T=1/|Q_{\rm tot, 22}|$ we
recover equation~(\ref{eq:eq11}) that is valid only for the pure array.

\subsubsection{Band oscillations}

To analyze the effects of the impurity on the band profiles it is
convenient, since $ T=|Q_{\rm tot, 22}|^{-1}$, to study in detail the
element $Q_{\rm tot,22}$ given by equation~(\ref{eq:eq17}). In the bands the eigenvalues
are complex, $\lambda_-=\lambda_+^*$ and of modulus one, so we can
(when considering the absolute value of $Q_{\rm tot,22}$) ignore from
our analysis the factor $\lambda_+^{N_1+N_2}$ in front of expression
(17). We now discuss the quantities $f_{\pm}$ and $\tilde Q_{ij}$.
The functions $f_\pm$ do not contain any information on the
impurity: they depend only on $\lambda_\pm$ and $Q_{11}$,
characterizing the periodic array. $f_{\pm}$ are real and smooth (
almost flat) functions within the bands, diverging as
$1/\sqrt{1-\xi^2}$ at the band edges \footnote{In the bands:
$\lambda_\pm =\xi \pm i \gamma$ with $\gamma=\sqrt{1-\xi^2}$
real, and $Q_{11}$ can be written as $Q_{11}=\xi +i \eta$. It follows from
the definition of $f_\pm$ that $f_- =-(\gamma +\eta)/2\gamma$ and
$f_+=f_-+1$. Given that $\eta$ and $\gamma$ are smooth functions everywhere
and $\gamma$ goes to zero at the band edges (where $\xi^2=1$) then $f_\pm$
are very smooth functions within the bands, diverging as $1/\sqrt{1-\xi^2}$
at the band edges.}. More important for our analysis is to
note that $f_+$ and $f_-$ take alternate roles in consecutive bands.
Namely, $|f_+|$ is larger than $|f_-|$ in the first and all
odd-numbered bands, while the opposite is true in the second and all
even-numbered bands.

The quantities $\tilde Q_{ij}$ are also typically very smooth
functions within the bands, diverging like $1/\sqrt{1-\xi^2}$ at the
band edges, and are all roughly of the same magnitude, except for
$\tilde Q_{21}$ which diverges at the Teflon resonance (but note
that $f_+=0$ at Teflon resonances).

\begin{figure}
\hspace*{2.5cm}\includegraphics[width=0.7\columnwidth, height=7cm]{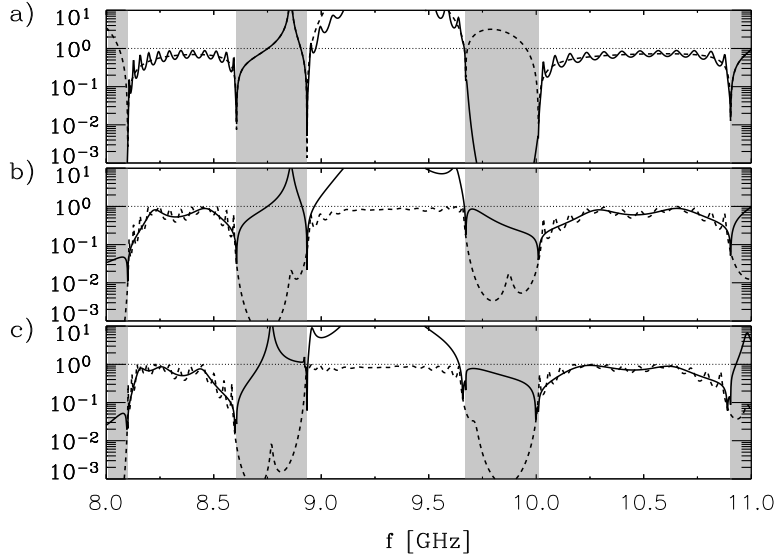}
\caption{\label{fig:fig6} Band oscillations due to impurities.
(a) and (b) correspond to an interstitial impurity {\it centered} in
the air spacing between the third and the fourth regular size Teflon
pieces, whereas in (c) it is {\it displaced}. $|f_+\tilde
Q_{11}|^{-1}$ (dashed) and $|f_+\tilde Q_{11}+f_+\tilde
Q_{21}\lambda_+^{-2N_2}|^{-1}$ (solid) are shown in (a). In (b) and
(c) $|\tilde Q_{\rm tot,22}|^{-1}$ (dashed) and $|f_+\tilde
Q_{11}-f_-\tilde Q_{12}\lambda_+^{-2N_1}|^{-1}$ (solid) are shown.}
\end{figure}

With this information about the $f_{\pm}$ and $\tilde Q_{ij}$ it is
expected that in the odd-numbered bands, where terms containing
$f_-$ can be ignored to first order, the first two terms in equation~(\ref{eq:eq17})
give the main contribution to $Q_{\rm tot,22}$. Specifically, $T
\approx|f_+(\tilde Q_{11}+\tilde Q_{21}\lambda_+^{-2N_2})|^{-1}$
should yield $N_2$ oscillations (recall that $\theta$ shifts through
$\pi$ in each band). Figure~\ref{fig:fig6}(a) shows this term and also $|f_+ \tilde
Q_{11}|^{-1}$ as a function of frequency for the case of an
interstitial impurity centered between the third and fourth Teflon
piece (the case corresponding to figure~\ref{fig:fig3}(b)). For the moment let us
concentrate on the first and third transmission band. The term $|f_+
\tilde Q_{11}|^{-1}$ (dashed curve in (a)) gives the zero order
approximation to the transmission band. According to the notation of
figure~\ref{fig:fig4} it follows that $N_1=2$ and $N_2=12$. The 12 oscillations can
be counted in figure~\ref{fig:fig4}(a) in the plot of $|f_+(\tilde Q_{11}+\tilde
Q_{21}\lambda_+^{-2N_2})|^{-1}$. Moreover, since $|f_+(\tilde
Q_{11}+\tilde Q_{21}\lambda_+^{-2N_2})|^{-1}$ gives the dominant
contribution in the odd-numbered bands, the $12$ oscillations should
also be observed in the total transmission. Indeed this is the case,
as it is shown by the dashed curve in figure~\ref{fig:fig6}(b). This plot also
shows that the total transmission curve oscillates about a slowly
oscillating curve (thick solid line) with two maxima. This slowly
oscillating curve is in fact given by $|f_+\tilde Q_{11} -f_-\tilde
Q_{12}\lambda_-^{-2N_1}|^{-1}$ with $N_1= 2$ as pointed out above.
Finally the last term in equation~(\ref{eq:eq17}) gives additional
$N_1+N_2$ oscillations of smaller amplitude, which provide small
corrections (not readily visible in the plot).

As figure~\ref{fig:fig6}(b) shows, the term $|f_+\tilde Q_{11} -f_-\tilde
Q_{12}\lambda_-^{-2N_1}|^{-1}$ does not agree at all with the
average curve of the transmission in the gaps nor in the {\it
{even}} numbered bands. This is because our discussion has been
limited to the bands, where the eigenvalues are complex, and of modulus
one, whereas in the gaps the eigenvalues are real, and hence the
factor $\lambda_+^{N_1+N_2}$ in (17) cannot be ignored (in the next
section we shall consider in detail the gap regions). There is no
agreement in the {\it {even}}-numbered bands either because there
the dominant terms are those containing the form factor $f_-$,
namely, the sum $|f_-(\tilde Q_{22}+ \tilde
Q_{12}\lambda_-^{-2N_2})|$. This can be readily seen by factoring in
equation~(\ref{eq:eq17}) the term $\lambda_-^{N_1+N_2}$ to get $Q_{\rm tot,22} =-
\lambda_-^{N_1+N_2}(f_-{\tilde Q_{22}}+ f_-\tilde
Q_{12}\lambda_-^{-2N_2}-f_+ \tilde Q_{21}\lambda_-^{-2N_1}-f_+
\tilde Q_{11}\lambda_-^{-2(N_1+N_2)})$. Written in this form, it is
clear that the same arguments used for the odd-numbered bands work
for the even numbered bands but with $f_-$ and $f_+$ interchanged.
Explicitly, the dominant term in the even-numbered bands is
$|f_-(\tilde Q_{22}+ \tilde Q_{21}\lambda_-^{-2N_2})|$, giving rise
again to $N_2=12$ oscillations. The second order term, producing the
$N_1=2$ oscillations, is $|f_-\tilde Q_{22} -f_+\tilde
Q_{21}\lambda_-^{-2N_1}|$.

Thus, {\it {the number of slow and fast oscillations appearing in the
profiles of the transmission bands can be read off directly from
equation~(\ref{eq:eq17})}}. Assuming for the moment that $N_2>N_1$, the band
profiles should show $N_2$ fast oscillations about a curve with
$N_1$ slow oscillations. This is exactly what we have noticed in our
experimental and numerical transmission curves, see, e.\,g.,
figure~\ref{fig:fig3}(b), corresponding to an interstitial impurity with
$(N_1,N_2)=(2,12)$, respectively.

On the other hand, if $N_1=N_2$, then the profile should show only
$N_1$ oscillations of larger amplitude than in the case above
($N_2\neq N_1$). The experiments confirm this, as it is illustrated
in figure~\ref{fig:fig3}(c), where $N_1=N_2=7$. Finally, if $N_1$ is close but not
exactly equal to $N_2$ then the oscillation pattern becomes more
complicated, with irregular (incomplete) oscillations and with no
distinctive modulation pattern, as exemplified by the plots of
figure~\ref{fig:fig5}, where $N_1$=6 and $N_2$=8.

Thus, given some experimental curves for the transmission, one can
determine where along the array there is a single impurity by simply
observing the pattern and number of oscillations. Note, however,
that due to time reversal symmetry it is not possible to deduce on
which side of the array to count the $N_1$ cells. At the very least,
the form of the profile indicates the presence of impurity or
defect, and whether it is located near the ends or near the center
of the array.

It is important to remark that the above conclusions hold as long as
the elements $\tilde Q_{ij}$ do not have oscillations of their own
as it is the case of the interstitial impurities discussed above.
For certain other types of defects the elements $\tilde Q_{ij}$ do
have some structure and consequently the type of profiles discussed
above become somewhat distorted. An extreme example occurs when a
smaller Teflon piece is placed off-center between two regular size
Teflon pieces, that is, a {\it {displaced}} interstitial impurity.
Specifically, $\tilde Q_{ij}$ show a strong frequency dependence
(e.\,g., $\tilde Q_{11}$ and $\tilde Q_{21}$ have each three extrema
in their real and imaginary parts in each band). For example,
figure~\ref{fig:fig6}(c) shows the total transmission and $|f_+\tilde Q_{11}
-f_-\tilde Q_{12}\lambda_-^{2N_1}|^{-1}$ for an interstitial
impurity placed off-center between the 3rd and 4th regular size
Teflon. This defect is described by the upper sketch of figure~\ref{fig:fig4} with
$N_1=2$ and $N_2=12$ with $N=16$. Note that the term $|f_+\tilde
Q_{11} -f_-\tilde Q_{12}\lambda_-^{2N_1}|^{-1}$ correctly gives the
average curve of the band profile but now there are three slow
oscillations and 14 fast oscillations, instead of 2 and 12
oscillations, respectively.

\subsubsection{Impurity states}

\begin{figure}
\hspace*{2.5cm}\includegraphics[width=0.7\columnwidth]{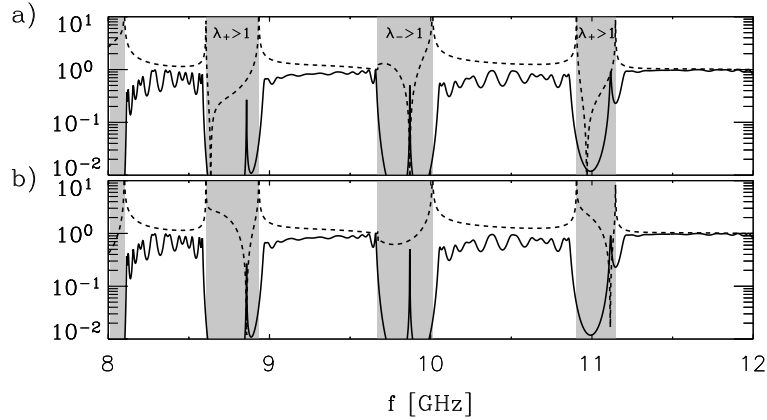}
\caption{\label{fig:fig7} Impurity states. In both plots the
transmission amplitude $1/|Q_{\rm tot,22}|$ is shown (solid). In (a) the
dashed lines corresponds to $\tilde Q_{22}$ and in (b) to $\tilde
Q_{11}$. All parameters are the same as in figure~\ref{fig:fig3}(c), except now the
impurity is between the 7th and 8th regular size Teflon piece, thus
$N_1=6$ and $N_2=8$. }
\end{figure}

We now analyze the appearance of the peaks in the gaps, the well
known signal for the presence of an impurity. Without limiting in
any way the conclusions to be drawn here, let us consider the
$d=d'=$4\,cm array, the example we have been using, and let us label
the gaps. Gap one is in the frequency range [7.7, 8.1]\,GHz, see,
e.\,g., figure~\ref{fig:fig7}; gap two in the range $[8.6, 8.9]$ GHz, and so on.

Recall that in the gaps the eigenvalues are real, therefore, we
rewrite equation~(\ref{eq:eq17}) as
\begin{equation}\label{eq:eq20}\fl
Q_{\rm tot,22} = f_+\tilde Q_{11}\lambda_+^N+f_+\tilde
Q_{21}\lambda_+^{N_1-N_2}-f_-\tilde Q_{12}\lambda_-^{N_1-N_2}
-f_-\tilde Q_{22}\lambda_-^N,
\end{equation}
where $N=N_1+N_2$. Now, for the transmission to produce a sharp peak
in the gaps, $Q_{\rm tot,22}$ must be small in a very narrow range of
frequencies. Note that $\lambda_+\lambda_-=1$, with $\lambda_-$ and
$\lambda_+$ alternatingly being the larger one in each gap. In the
even-numbered gaps (where $\lambda_+>1>\lambda_-$) the dominant term
in $Q_{\rm tot,22}$ is the first term in equation~(\ref{eq:eq20}), and it will be large
(corresponding to a negligible transmission) unless it happens that
$\tilde Q_{11}$ is very small or zero. Similarly, in the
odd-numbered gaps (where $\lambda_->1>\lambda_+$), the dominant term
is the last one and it will also be a large number unless $\tilde
Q_{22}$ happens to be very small. To illustrate this general
observation, in figure~\ref{fig:fig7} we consider the case of an interstitial
impurity placed in the air spacing of the 8th cell in a 16-cell
array (thus, $N_1=6, N_2=8$). In figure~\ref{fig:fig7}(a) we show the transmission
amplitude $1/|Q_{\rm tot,22}|$ and also the term $|\tilde Q_{22}|$. In
contrast, in (b) the term $|\tilde Q_{11}|$ is shown, together with
$1/|Q_{\rm tot,22}|$. Observe in figure~\ref{fig:fig7}(a) that whenever $\tilde Q_{22}$
goes sharply to zero in the odd-numbered gaps ($\lambda_->1$), there
is an impurity state at the same frequency. $\tilde Q_{22}$ also
goes to zero sharply in the even-numbered gaps($\lambda_+>1$) but
here there is no coincidence with the impurity states, in agreement
with our argument of the previous paragraph. Similarly, figure~\ref{fig:fig7}(b)
shows that there is an impurity state in the even-numbered gaps at
the frequency values where $\tilde Q_{11}$ goes to zero. Thus, our
simple argument above really works in determining the position of
the impurity states: {\it {an impurity state occurs in odd-numbered
gaps at the frequencies where $\tilde Q_{22}$ is zero and in the
even-numbered gaps where $\tilde Q_{11}$ is zero}}. A similar case
was already shown in figure~\ref{fig:fig3}(c) except there ($N_1=N_2=7$ and
thus band profiles are different but peak positions are the same),
where we see an excellent agreement with experiment concerning the
location of the peaks.

Single impurity states are known to decay exponentially away from
the site of the impurity. One then may expect that the intensity of
the peak should be stronger the closer the impurity is from any of
the ends of the array. However close inspection of formula $(20)$
indicates that this is not the case. That is, exactly at the
impurity state frequency, say when $\tilde Q_{11}$ is zero in an
even-numbered band, the first term in (20) is zero and the leading
terms are $f_+\tilde Q_{21}\lambda_+^{N_2-N_1}-f_-\tilde Q_{12}
\lambda_-^{N_2-N_1}$. The inverse of its absolute value gives to a
good approximation the intensity of the peak. For simplicity, let us
consider first the case $f_+\tilde Q_{21}=f_-\tilde Q_{12}$ which is
valid only for ``symmetric impurities'', i.\,e., when the distance
$a$ equals the distance $c$ (see figure~\ref{fig:fig4}). Note that
$\lambda_+^{N_2-N_1}+\lambda_-^{N_2-N_1}$ as a function of $N_2-N_1$
is symmetric about its minimum $N_2-N_1=0$. Thus, recalling that the
transmission is the reciprocal of $|Q_{\rm tot,22}|$, we conclude that
{\it {the intensity of the impurity state should be stronger the
closer the impurity is to the center of the array}}. It is strongest
when $N_2=N_1$, i.\,e., when the impurity or defect is at the center
of the array.

The above argument assumed that the impurity is symmetric. If it is
not the case, then it can be shown that the minimum of $|f_+\tilde
Q_{21}\lambda_+^{N_2-N_1}-f_-\tilde Q_{12} \lambda_-^{N_2-N_1}|$
does not occur when the impurity is at the center ($N_1=N_2$) but at
a distance $N_1-N_2$ that is proportional to the logarithm of
$|f_-\tilde Q_{12}|/|f_+\tilde Q_{21}|$. Since this ratio is not
large, the highest intensity peak occurs still when the impurity is
close to the center of the array. This was verified in all our
experiments and can be seen clearly by comparing figure~\ref{fig:fig3}(b) and
figure~\ref{fig:fig3}(c). We emphasize that this effect is independent of the type
of the periodic potential and type of impurity in a 1D periodic
array. In fact, the same behavior was noted in an experiment with a
coaxial connector photonic crystal \cite{Pra99} where, however, no
explanation was given.

As we have seen, the intensity of the peaks always decreases as the
number $N$ of cells increases. However, an important conclusion,
drawn from a detailed examination of equation~(\ref{eq:eq20}) as a function of
$\lambda_+$, for fixed $N$, is the following: {\it {the decrement in
the peak intensity as $N$ increases is weaker the farther the
impurity state is from the center of the transmission band}}. This
was confirmed by our experimental and theoretical results and it is
in agreement with the generally accepted idea that shallow states
near the band edges are the shallow levels, associated with long
range potentials, in contrast with the deep levels lying near the
center of the band (see, e.g.,\cite{Men99}).

\section{Conclusions}

We analyzed the effects on the transmission of single impurities in
an array of regularly spaced pieces of Teflon in a microwave guide,
described by a 1D photonic Kronig-Penney Model. We performed a
series of experiments with point defects of various types; namely,
interstitial, substitutional, and vacancies and showed that single
impurity affects the transmission in two ways. One is the well known
appearance of localized states in the gaps; the second, to our
knowledge not discussed so far in the literature, is the appearance
of fast and slow oscillations in the band profiles. Our transfer
matrix calculations correctly predicted these features. Experiments
with other types of single impurities, e.\,g., displaced
interstitial and displaced substitutional impurities, not reported
here, were also found to be equally well described by our model. The
transfer matrix calculations (being purely numerical) agree very
well with the experimental data, but they provide no insight for the
understanding of the various features observed. Thus, as an
important contribution, we derived an exact closed form expression
for the transmission amplitude that is useful in elucidating the
effects of a single impurity on the band profiles, the impurity
levels, and their intensity. This expression involves elements of
the transfer matrix $Q$ of the regular cells in the array and of the
transfer matrix $\tilde Q$ of the impurity. Since this matrix is
written in the representation that diagonalize the transfer matrix
of the regular cells, $\tilde Q$ contains information about the
impurity and its coupling with the host environment. We found that
the set of impurity levels is given by the zeroes of the diagonal
elements of $\tilde Q$. Further it was shown that the intensity of
the impurity states in the gaps depends on two factors, namely, the
off-diagonal elements of $\tilde Q$ and the distance of the impurity
from the center of the array. The closer the impurity is to any of
the ends of the array, the lower the intensity of its level, and
vice versa. It was also shown that the number of fast and slow
oscillations in the bands give direct information about the position
of the impurity, relative to the center of the array.

We note that the agreement between the experimental results and
analytical calculations gives us confidence to treat the inverse
problem. That is, we can extract information about the unknown
defect from the inspection of the band profiles and the localized
defect modes. Our method is an alternative approach to the
tight-binding and Green's functions methods, for 1D systems, with
the advantage that it is simpler and elucidating. Although the
experimental realization here is in the microwave regime, the model
and the results are equally valid for higher scales of frequency
corresponding to light experiments. We remark  that we can apply our
formalism in a straight-forward manner to any kind of regular array
with single defects once we have determined the particular transfer
matrices for the regular cell and for the impurity.

We emphasize that the formulas (17-20) and the procedure for
determining the position and intensity of impurity states and
features of the band profiles for the photonic Kronig-Penney model
are exactly the same as for the {\it electronic} Kronig-Penney model
(with square barriers) for energies above the barrier. This is so
since their transfer matrix elements are identical and the
difference is only in the definition of the longitudinal wave
vector. Hence our procedure is useful for analyzing the effects of
impurities either, in the transport of charged particles in 1D
periodic arrays of electronic potentials (e.\,g., heterostructures)
or the transmission of electromagnetic waves in 1D photonic arrays.

Finally, the fact that our experimental and theoretical results
agree very well give us confidence to treat various other types of
specific arrangements with the goal of realizing them in other
experimental set-ups as photonic devices.
We would like to stress that this is by no means self evident.
The observed transmission patterns are the result of a
complicated interplay of interferences from all structures
within the waveguide, and it is well known that interferences
are extremely sensitive to perturbations, in particular in the
presence of absorption. The present work has shown that we can rely
on the experiment in this respect. Thus our experimental
set-up can be used as a preliminary --- and low cost --- study of a
particular Kronig-Penney model to be implemented later in more
sophisticated and expensive microscopic photonic realizations, which
might have real applications.

\section*{Acknowledgments}
One of the authors (G.A.L-A) is grateful to the DFG, Germany, for
the economic support via the Mercator Professorship and to
Prof.~St\"{o}ckmann and his group for their hospitality. The
experiments were supported by the DFG. G.~A.~L-A also acknowledges
partial support from SEP-CONACYT Convenio P51458.

\end{document}